\documentclass[authoryear]{elsarticle}

\usepackage[T1]{fontenc}
\usepackage[british]{babel}
\usepackage[utf8]{inputenc}
\usepackage[caption=false]{subfig}
\usepackage{mathptmx}
\usepackage{amsmath}
\usepackage{tikz}
\usepackage{color}
\usepackage{url}
\usepackage{graphicx}
\usepackage{cleveref}
\usepackage{todonotes}
\usepackage{booktabs}   
\usepackage{xspace}
\usepackage{stmaryrd}
\usepackage[noprefix]{nomencl}	
\usepackage{lmodern}
\usepackage{natbib}
\usepackage{epstopdf}
\usepackage{upgreek}
\usepackage{todonotes}
\usepackage[caption=false]{subfig}

\crefname{chapter}{Chapter}{Chapters}
\crefname{section}{Section}{Sections}
\crefname{appendix}{Appendix}{Appendices}
\crefname{subsection}{Section}{Sections}
\crefname{subsubsection}{Section}{Sections}
\crefname{equation}{Equation}{Equations}
\crefname{figure}{Figure}{Figures}
\crefname{table}{Table}{Tables}
\crefname{subfigure}{Figure}{Figures}
\crefname{listing}{Listing}{Listings}

\usepackage[activate={true,nocompatibility},final,kerning=true,tracking=true,spacing=true]{microtype}
\microtypecontext{spacing=nonfrench}

\newcommand{\eg}{e.g.\xspace}
\newcommand{\ie}{i.e.\xspace}

\renewcommand{\v}[1]{\ensuremath{\mathbf{#1}}} 
\newcommand{\gv}[1]{\ensuremath{\mbox{\boldmath$ #1 $}}}  
\renewcommand{\d}[2]{\frac{d #1}{d #2}} 
\newcommand{\pd}[2]{\frac{\partial #1}{\partial #2}} 
\newcommand{\grad}[1]{\gv{\nabla} #1} 
\renewcommand{\div}[1]{\gv{\nabla} \cdot #1} 
\let\baraccent=\= 
\renewcommand{\=}[1]{\stackrel{#1}{=}} 
\newcommand*{\unit}[1]{\ensuremath{\,\mathrm{#1}}}
\newcommand{\nomunit}[1]{\renewcommand{\nomentryend}{\hspace*{\fill}#1}}

\newcommand{\we}{\operatorname{\mathit{W\kern-.24em e}}}
\newcommand{\re}{\operatorname{\mathit{R\kern-.08em e}}}
\newcommand{\eo}{\operatorname{\mathit{E\kern-.08em o}}}

\makenomenclature

\title{The transition in settling velocity of surfactant-covered
droplets from the Stokes to the Hadamard-Rybczynski solution}

\date{\today}
\author[asmund1,asmund2]{Åsmund Ervik\corref{cor1}}
\ead{asmunder@pvv.org}
\author[erik]{Erik Bjørklund}
\address[asmund1]{Department of Energy and Process Engineering, Norwegian University of
Science and Technology (NTNU), NO-7491 Trondheim, Norway}
\address[asmund2]{SINTEF Energy Research, P.O. Box 4761, Sluppen, NO-7465 Trondheim, Norway}
\address[erik]{Sulzer Chemtech, Solbråveien 10, NO-1383 Asker, Norway}

\begin{document}

\begin{abstract}
The exact solution for a small falling drop is a classical result by Hadamard
and Rybczynski. But experiments show that small drops fall slower than 
predicted, giving closer agreement with Stokes' result for a falling
hard sphere. Increasing the drop size,
a transition between these two extremes is found. This is due to
surfactants present in the system, and previous work has led to the stagnant-cap
model. We present here an alternative approach which we call the
continuous-interface model. In contrast to the
stagnant-cap model, we do not consider a surfactant advection-diffusion
equation at the interface. Taking instead the normal and tangential
interfacial stresses into account, we solve the Stokes equation analytically
for the falling drop with varying interfacial tension. Some of the solutions
thus obtained, \eg the hovering drop, violate conservation of energy unless
energy is provided directly to the interface. Considering the energy budget
of the drop, we show that the terminal velocity is bounded by the Stokes and
the Hadamard-Rybczynski results. The
continuous-interface model is then obtained from the force balance 
for surfactants at the interface. The resulting expressions gives the transition
between the two extremes, and 
also predicts that the critical radius, below which drops fall like hard
spheres, is proportional to the \emph{interfacial} surfactant
concentration. By analysing experimental results from the
literature, we confirm this prediction, thus providing strong arguments for the
validity of the proposed model.
\end{abstract}


\maketitle

\printnomenclature[1.5cm]

\section{Introduction} 
A single falling drop is one of the simplest two-phase flow
configurations, and has been under scrutiny since the dawn of fluid mechanics
research. Many of the early studies were focused on drops impacting a pool of
water, such as the works by \citet{worthington1876} and by \citet{reynolds1875}.
\citet{stokes1851} was the first
to give an analytical solution for the flow at low Reynolds number ($\re$) around
a solid sphere falling at terminal velocity. Then
\citet{hadamard1911} and \citet{rybzynski1911} independently
published the analytical solution for the flow inside and around a clean
spherical drop falling at low $\re$. This has later been extended by various authors to
account for the presence of surfactants, under various assumptions, as will be
discussed in the following.

The case of liquids with surfactants may seem to be of lesser interest than the case of
clean fluids. But experimentally observed terminal velocities of small drops do
not match the Hadamard-Rybczynski result, but rather the Stokes result for the
vast majority of combinations of ``clean'' fluids, see \eg the work
by \citet{nordlund1913,lebedev1916,silvey1916,bond1927,bond1928}. In the latter
work, a distinguished jump was found in the terminal velocity, going from the Stokes result to the
Hadamard-Rybczynski result as the drop radius was increased. This has been
confirmed in later experiments, \eg by \citet{griffith1962}.

It is noteworthy that even Hadamard acknowledges the fact that his expression
does not agree with experimental results, in the
closing words of his 1911 paper, where he refers to disagreement between the
expression and some (at that time) unpublished experimental results:
\begin{quote}
  \emph{
La formule (III) présente, avec les résultats expérimentaux obtenus quant à présent (et encore inédits), de notables divergences. Il semble donc, jusqu'á nouvel ordre, que, dans les cas étudiés, les hypothèses classiques dont nous sommes parti doivent être modifiées.
}
\end{quote}

In fact there are extremely few published
works that are able to obtain terminal velocities for very small drops
matching the Hadamard-Rybczynski result, and then only for quite singular fluid
combinations. Examples include molten lead drops in liquid beryllium trioxide
\citep{volarovich1939}, or liquid mercury
drops in highly purified glycerine \citep{frumkin1947a}. There are a few studies where the
authors have gone to great pains to purify more ordinary fluid systems, but these have
been limited to $\re > 10$, see \eg \citet{thorsen1968,edge1972}. That one is
able to obtain agreement with the Hadamard-Rybczynski result for small drops only when at least
one of the fluids in question are chemically quite different from ordinary
liquids, supports the hypothesis that amphiphilic surfactants, which occur naturally in even
highly purified organic liquids, are the cause of this phenomenon.

In later years, attention towards surfactants and their role in systems both
man-made (\eg in various foods) and natural (\eg in our lungs) has increased
considerably. As an example, it is recognised that surfactants play a dominant
role in the stability of emulsions \citep{lucassen1996}, whether this 
stability is desired (as in mayonnaise) or not (as in a water-crude oil
emulsion). Surfactants act both to slow down the sedimentation of drops, and to 
prevent the coalescence of drops in a separation process.

Several authors have considered the extension of the Hadamard-Rybczynski
analytical result
to account for the presence of surfactants. Prominent
examples include the work by \citet{frumkin1947},
\citet{savic1953}\footnote{The report by Savic has not been available electronically in the past;
  however we were informed by the National Research Council of Canada that the
  copyright on it has expired, and have thus made a scanned copy available at
  \texttt{http://archive.org/download/mt-22/savic.pdf}},
  \citet{davis1966}, \citet{griffith1962} and \citet{sadhal1983}. This body of work
incorporates both experimental data, in the form of correlations, and exact
results under various assumptions; see \citet[Chapter II.D]{clift1978} for
a review.

\begin{figure}
   \begin{center}
     \includegraphics[width=0.8\linewidth]{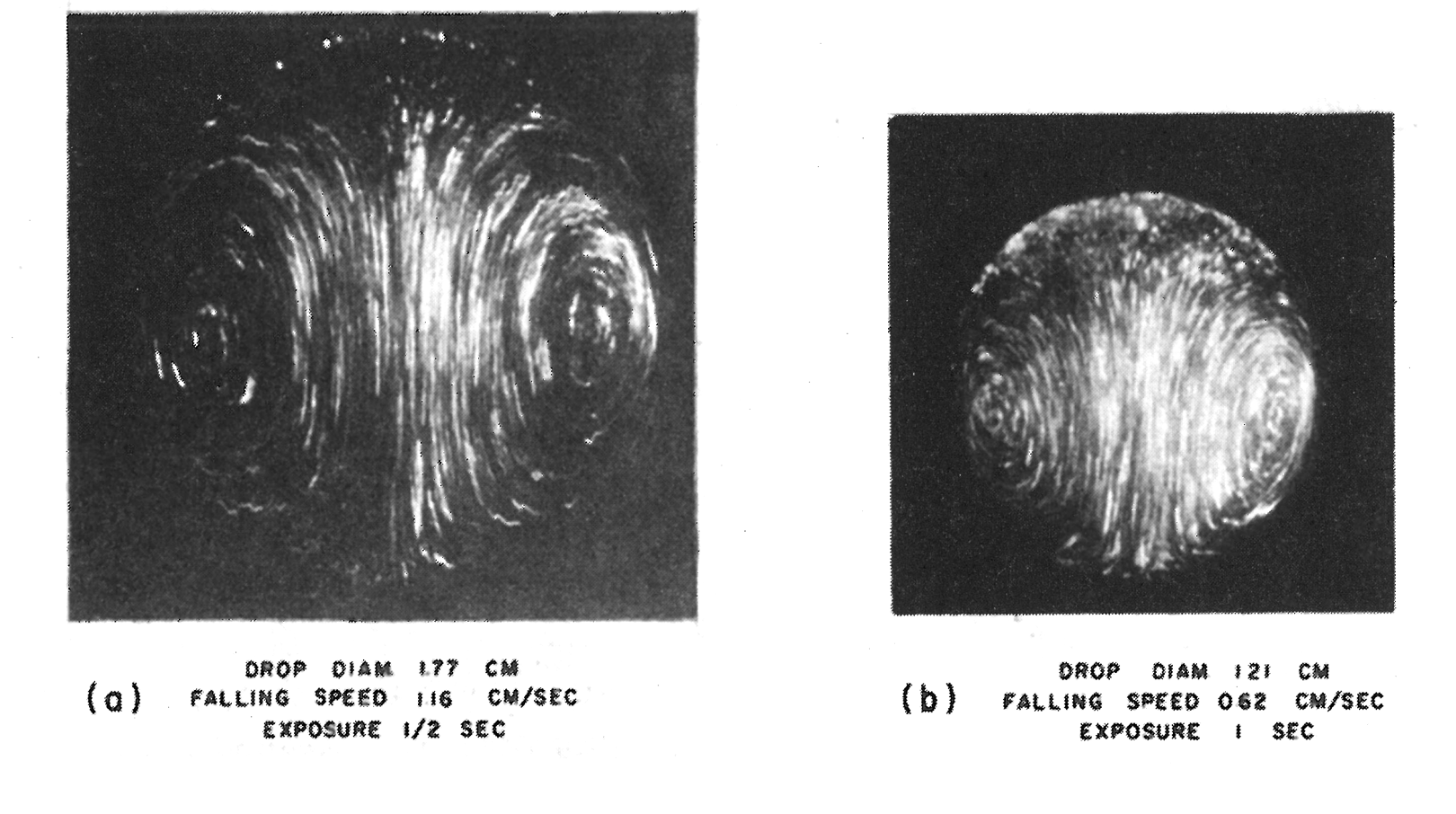}
   \end{center}
   \caption{From the flow visualisation studies by Savic.
     Image showing a falling water
     drop in castor oil, with complete internal circulation, (a), and a stagnant
     cap, \ie a downwards shift of the internal flow pattern, (b). 
     Reproduced from \citet{savic1953} with permission from the
     National Research Council of Canada.}
   \label{fig:savic}
 \end{figure}

A prominent feature in these works is the assumption of a
stagnant cap, \ie that the surfactant is accumulated at the top of a falling drop,
such that the interface is immobile in this region and free to move on the rest
of the drop. This assumption is based on photographic evidence gathered for
larger drops. An example is the photograph in the paper by \citet{savic1953},
reproduced here in \cref{fig:savic}, which is often taken
as \emph{prima facie} evidence for the stagnant cap model.

In this paper, we will extend the derivation by \citet{chang1985} of the analytical solution for the
terminal velocity of a low $\re$ circular drop with an arbitrary surfactant
(hence interfacial tension) distribution, taking here also the normal interfacial
stresses into account. 
We show that this leads to a plethora of solutions, some of which are clearly
unphysical (in the absence of an external energy input), such as a hovering drop. 
By appealing to the conservation of energy,
we show that the physically admissible terminal velocities are bounded from below by the
Stokes result for hard spheres, and from above by the Hadamard-Rybczynski result.

We proceed to supplement this with a simple model for the forces
acting on surfactant molecules at the interface, giving an expression for the
transition in terminal velocity between the two extremal values. This expression
depends on the properties of the surfactant in question. From the theory we
predict that for a given surfactant, the critical radius $R_c$ below which drops fall like hard
spheres, should be proportional to the \emph{interfacial} surfactant concentration. To
confirm this prediction we determine the critical radii for the different \emph{bulk}
surfactant concentrations considered in the experiments performed by
\citet{griffith1962}. We demonstrate that these critical radii, when plotted
against the \emph{bulk} surfactant concentration, all collapse on a single
Langmuir isotherm. Since the Langmuir isotherm relates the interfacial and the 
bulk surfactant concentration, this confirms the prediction by the present
model, which we call the continuous-interface model.

We also discuss briefly the existing versions of the
stagnant-cap model and to compare these with the model presented in this work.
One should note that the straight-forward application of the stagnant-cap
model gives rise to certain pecularities.
As an example, consider Equation 7-270 on page 497 in the book by
\citet{leal2007}, which reads 
\begin{equation}
  \grad_s \cdot (\v{u}_s\Gamma) \approx 0
\end{equation}
where $\Gamma$ is the surfactant concentration.
This equation follows from the assumption that
the surfactant is insoluble, and that the interfacial
P{\'e}clet number is $Pe_s = 2 R |\v{u}_s|/D_s \gg 1$ where 
$\v{u}_s$ is the velocity at the interface and $D_s$ is the interfacial
diffusivity of the surfactant. 
From this, the classic stagnant cap result is obtained, namely that the part of the
interface where the surfactant is found has $\v{u}_s = 0$ (this is the stagnant
cap), and the rest of the interface has $\Gamma = 0$. But if $\v{u}_s = 0$ in the
stagnant cap region where the surfactant is located, then $Pe_s = 0 \not\gg 1$.
Thus one is lead to consider what velocity is the appropriate to use in the
surface P{\'e}clet number, if one is to keep the $Pe \gg 1$ assumption.

It should be noted that while the result obtained in this
work for the interfacial tension distribution along the drop interface has the same
functional form as the result obtained in the classic analysis \eg by
\citet{levich1962}, unlike Levich, we do not assume the variation in
surfactant concentration to be small, and since the present work avoids the use of
a surfactant advection-diffusion equation on the interface (in contrast to
previous approaches) we are able to obtain simultaneous analytical solutions to
the flow and the interfacial tension distribution. This has been a major obstacle in
previous work, as noted \eg by \citet{leal2007}:
\begin{quote}
\emph{It is not generally possible to obtain analytic solutions of the resulting
problem because of the complexity of the surfactant transport phenomenon and the
coupling between surfactant transport and fluid motion.}
\end{quote}
In closing, we argue that the present model could be more appropriate for interfacially active
agents which are amphiphilic molecules, while the stagnant-cap model may be more
appropriate for dispersed microscopic particulates which adsorb at the interface and
thus modify the boundary conditions of the problem.

\section{Theoretical results}
\subsection{Governing equations}
\label{sec:theory}
The flow field $\v{u}$ of an incompressible viscous Newtonian fluid is governed
by the Navier-Stokes equations on the form
\begin{align}
  \div\v{u} &= 0 \label{eq:ns-divfree} \rm{,} \\
	\pd{\v{u}}{t}+(\v{u}\cdot\grad)\v{u} &= - \frac{\grad{p}}{\rho}
  + \frac{\mu}{\rho}\grad^2\v{u} + \v{f}_\text{b} \rm{.} \label{eq:ns}
\end{align}
Here $p(\v{x})$ is the pressure field and $\v{f}_\text{b}$ is some external acceleration, such as
gravity. As it stands, this system of equations is closed when fluid properties
and initial and boundary conditions are given. 
\nomenclature{$\mu$}{Dynamic viscosity of a fluid.\nomunit{Pa$\cdot$s}}%
\nomenclature{$\rho$}{Density of a fluid.\nomunit{kg/m$^{3}$}}%
\nomenclature{$\v{f}_\text{b}$}{External acceleration.\nomunit{m/s$^2$}}%

The system can be extended to 
two fluids by specifying an interface that
separates fluid $_1$ with properties $\rho_1, \mu_1$ from fluid $_2$ with
properties $\rho_2,\mu_2$, as well as two dynamic interfacial relations related to the interfacial tension $\sigma$. For the
case of a drop or bubble we will mark the internal properties with
$_1$ and the external properties with $_2$.

In order to have closure of \cref{eq:ns} with this extension, one also needs 
the following interfacial relations for two-phase flow:
\begin{align}
  \label{eq:ujump}
    \llbracket\v{u} \rrbracket =& \v{0},\\
    \v{t}\cdot\llbracket\v{T}\rrbracket\cdot\v{n}=&-\v{t}\cdot\nabla
    \sigma,\label{eq:marangoni}\\
    \v{n}\cdot\llbracket\v{T}\rrbracket\cdot\v{n}=&\kappa \sigma.\label{eq:laplace-young}
\end{align}
\nomenclature{$\v{T}$}{Cauchy stress tensor. \nomunit{N/m$^2$}}%
\nomenclature{$R$}{Radius of a drop.\nomunit{m}}%
\nomenclature{$D$}{Diameter of a drop.\nomunit{m}}%
\nomenclature{$\v{g}$}{Gravitational acceleration.\nomunit{m/s$^2$}}%
\nomenclature{$\sigma$}{Interfacial tension.\nomunit{N/m}}%
\nomenclature{$\kappa$}{Interfacial curvature.\nomunit{1/m}}%
Here the jump in a quantity across the interface is denoted by
$\llbracket$-$\rrbracket$, $\v{n}$ is the normal and $\v{t}$ is a unit vector in
the tangent plane to the interface, and $\v{T}$ is the stress tensor. We choose the normal vector to
point out from a drop, and the jump is then given by \eg $\llbracket \mu
\rrbracket = \mu_2 - \mu_1$, \ie the difference between the bulk and the
drop value. In the case of a spherical droplet with zero velocity field,
\cref{eq:laplace-young} reduces to the Young-Laplace relation for the pressure
difference across an interface ($\Delta p = 2\sigma/R$). The Marangoni force
comes in through \cref{eq:marangoni}, and the functional form of the
coefficient of interfacial tension along the drop interface, $\sigma$, is also
needed for closure.

In this paper, the case of one spherical droplet falling in an unbounded domain
will be considered. In this case, it is natural to
introduce the following characteristic properties:
\begin{equation}
x^\ast=R,\quad u^\ast=U, \quad t^\ast=R/U, \quad p^\ast=\sigma/R,
\end{equation}
\nomenclature{$U$}{Free stream velocity.\nomunit{m/s}}%
giving the following non-dimensional Navier-Stokes equations:
\begin{align}
  \div\v{u} &= 0 \label{eq:non-ns-divfree} \rm{,} \\ \pd{\v{u}}{t}+(\v{u}\cdot\grad)\v{u}&=\frac{1}{\re}\left[
\grad^2\v{u}+\frac{\re}{\we}\left(\eo\,\v{f}-\grad{p}\right)\right] \rm{.}\label{eq:non-ns}  
\end{align}
Here, $\re$ denotes the Reynolds number, $\we$ is the Weber number and $\eo$ represents the Eötvös number. The Reynolds number is given by 
$\re = \rho_2 U R/\mu_2$ as it is customary to use the continuous fluid properties in the dimensionless groups. This dimensionless number gives the ratio of
inertial forces to viscous forces. The Weber number is given by 
$\we = \rho_2 U^2 R/\sigma$ and gives the ratio between inertial forces and
interfacial tension forces. Lastly, the Eötvös number, $\eo=\rho_2 g R^2/\sigma$,
gives the ratio between the body forces and the capillary forces.

In what follows, $\re$, $\we$ and $\eo$ are all assumed to be small. The
assumption of $\re\ll 1$ and of steady state flow simplifies the Navier-Stokes 
equation to the steady Stokes equation:
\begin{equation}
  \grad^2\v{u}+\frac{\re}{\we}\left(\eo\,\v{f}-\grad{p}\right)=\v{0}\rm{.}
  \label{eq:stokes}
\end{equation}

When $\we$ is small, $\we \ll 1$, the forces due to interfacial tension determine the
shape of the interface through minimising the interfacial energy, which results in
a spherical drop. As demonstrated by the Hadamard-Rybczynski result, the
spherical falling drop is an exact solution to the Stokes equation. Furthermore, as
demonstrated by \citet{kojima1984} and in subsequent work (see \citet[Chapter
6]{stone1994} for a review), perturbations away from the spherical shape for
a drop falling at low $Re$ will either relax back toward the spherical shape,
or form instabilities as elongated tails.

Lastly, the assumption of a small Eötvös number means that the body forces 
are small compared to the capillary forces, and hence will not alter the spherical
shape of the droplet, but rather induce an acceleration on the droplet as
a rigid body. In the case without surfactants, \citet{taylor1964} considered the deviation from
a spherical shape at low $\eo$ and found that this is $O(\re^2)$, \ie very small when $\re$
is small. In general, the assumption of a spherical drop is not a significant
restriction, and so this assumption is
ubiquitous in the literature on the falling drop at low $\re$ with 
surfactants \citep{savic1953,levich1962,griffith1962,davis1966,sadhal1983,leal2007}.

\subsection{Spherical droplet in a quiescent liquid}\label{sec:sphere_drop}
We will now consider a spherical droplet in a gravitational field surrounded by a quiescent
liquid with which the droplet is immiscible. For a perfectly clean
interface, the stationary solution is given by the Hadamard-Rybczynski
solution. We proceed to let the interfacial tension vary along the
interface and investigate the solutions obtained when accounting for
the Marangoni force. We will follow in the steps of the analysis of
\citet{chang1985}, but we will also include the interfacial
conditions for normal stresses. The appropriate boundary conditions
are then given by \cref{eq:ujump,eq:marangoni,eq:laplace-young}.

\begin{figure}[htbp]
  \centering
  \includegraphics[width=0.35\linewidth]{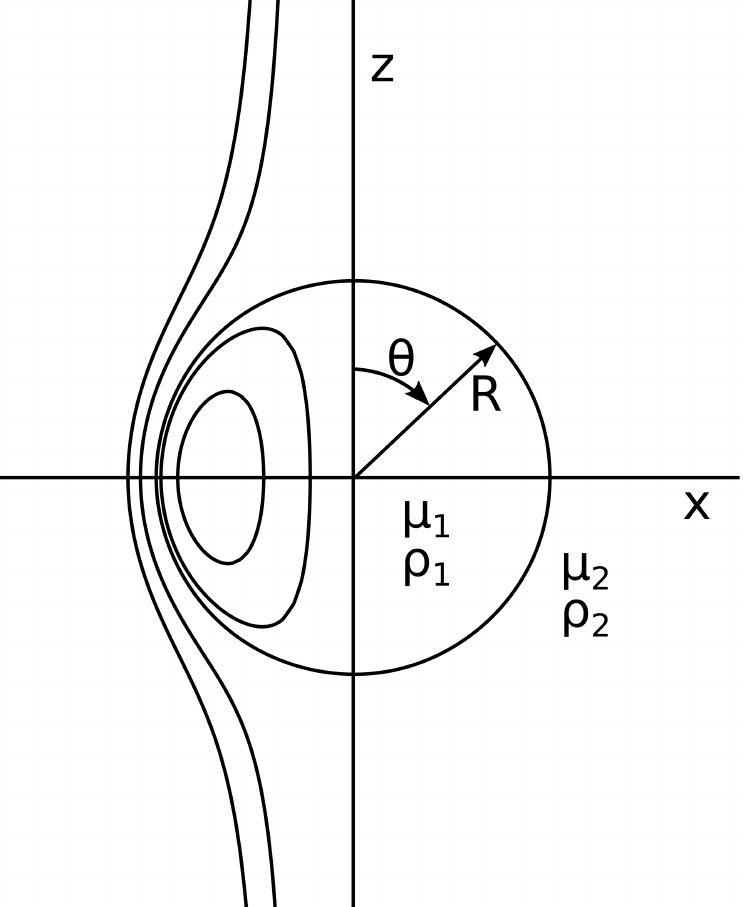}
  \caption{Illustration of coordinate system adopted, showing also the labelling
  of drop/bulk fluid properties.}
  \label{fig:coordinates}
\end{figure}

We employ a spherical coordinate system $(r,\theta,\phi)$ fixed at the centre-of-mass of the
droplet, with polar angle $\theta$ measured from the positive z-axis. For
convenience we will at times refer to the axes of the Cartesian coordinate
system, which has positive x-axis corresponding to $\theta = \pi/2, \phi = 0$.
This is illustrated in \cref{fig:coordinates}.
The situation is cylindrically symmetric, so the azimuthal angle $\phi$ is
redundant. The normal vectors in the $r$ and $\theta$ directions are denoted by
$\v{e}_r$ and $\v{e}_\theta$, respectively.
For a spherical drop, the normal (\ie radial) velocity is zero at the interface, and the
velocity far away from the droplet is given by
\begin{equation}
  \v{u}=U\cos\left(\theta\right)\v{e}_{r}-U\sin\left(\theta\right)\v{e}_{\theta},
\end{equation}
where $U$ is the uniform velocity at
infinity. The general solution for the stream functions outside 
and inside the droplet are found to be
\nomenclature{$C^{-1/2}_{n}$}{$n$'th Gegenbauer polynomial. \nomunit{1}}%
\nomenclature{$r$}{Radial distance from droplet centre. \nomunit{m}}%
\nomenclature{$\theta$}{Polar angle, measured from the z-axis. \nomunit{rad}}%
\begin{align}
    \Psi_{2} =&\sum_{n=2}^{\infty}
    \left(A_{n}r^{n}+ B_{n}r^{1-n}+D_{n}r^{2+n}+E_{n}r^{3-n}
    \right)C^{-1/2}_{n}(\theta),\\
    \Psi_{1} =&\sum_{n=2}^{\infty}
    \left(F_{n}r^{n} +G_{n}r^{1-n}+ H_{n}r^{2+n}+I_{n}r^{3-n}
    \right)C^{-1/2}_{n}(\theta),
\end{align}
where the velocity components are related to the stream function by
\begin{align}
u_{r}=\frac{1}{r^2 \sin\left(\theta\right)}\pd{\Psi}{\theta},\\
u_\theta=-\frac{1}{r\sin\left(\theta\right)}\pd{\Psi}{r}.
\end{align}
Requiring a uniform velocity at infinity gives
\begin{align}
    A_{2}&=- U,\\
    A_{n}&=0\quad \forall n \geq 3,\\
    D_{n}&=0\quad \forall n\geq 2,
\end{align}
and the assumption that the velocity is bounded at the origin gives
\begin{align}
    G_{n}&=0 \quad\forall n\geq 2,\\
    I_{n}&=0 \quad\forall n\geq 2.
\end{align}
The vanishing normal velocity at the interface ($r=R$) gives
\begin{align}
    B_{2}&=R^{3} U-R^{2}E_{2},\\
    B_{n}&=-R^{2}E_{n} \quad\forall n\geq 3,\\
    H_{n}&=-\frac{F_{n}}{R^{2}}\quad\forall n\geq 2.
\end{align}
The final kinematic boundary condition, the continuity of the velocity
field across the interface, gives
\begin{align}
    E_{2}&=\frac{3}{2}R U-RF_{2},\\
    E_{n}&=-R^{2n-3}F_{n} \quad\forall n\geq 3.
\end{align}
Both stream functions can now be expressed through one common set of
coefficients
\begin{align}
    \Psi_{2} =&
    \left(- Ur^{2}-\left(\frac{R^{3}}{2} U-R^{3}F_{2}\right)\frac{1}{r}
          +\left(\frac{3}{2}R U-RF_{2}\right)r\right)C^{-1/2}_{2}\notag\\
    +&\sum_{n=3}^{\infty}\left(\frac{R^{2n-1}}{r^{n-1}}-\frac{R^{2n-3}}{r^{n-3}}\right)F_{n} C^{-1/2}_{n},\\
    \Psi_{1} =&\sum_{n=2}^{\infty}
    \left(r^{n}-\frac{r^{n+2}}{R^{2}}\right)F_{n}C^{-1/2}_{n},
\end{align}
where $R$ is the radius of the droplet.

The dynamic interfacial conditions will now be used to determine the
last coefficient and the interfacial tension as a function of the polar
angle. Since the Legendre polynomials form a complete orthonormal
basis for any periodic function, we may write
\begin{equation}\label{eq:sigma_legendre}
  \sigma=\sum_{n=0}^{\infty}\sigma_{n}\text{P}_{n}\left(\eta\right),
\end{equation}
where $\eta=\cos\left(\theta\right)$ and $\text{P}_{n}$ is the $n$'th Legendre polynomial. The normal
stress condition (\cref{eq:laplace-young}) in spherical coordinates can be written as
\nomenclature{$\text{P}_n$}{$n$'th Legendre polynomial. \nomunit{1}}%
\begin{equation}
 \llbracket-p\rrbracket+2\llbracket\mu\frac{\partial}{\partial r} 
 \left(\frac{1}{r^{2}}\pd{\Psi}{\eta}\right)\rrbracket=\frac{2}{R}\sigma,
\end{equation}
giving the following relations between $F_{n}$ and $\sigma_{n}$
\begin{align}
  \sigma_{0}=&\frac{R}{2}\left(p_{01}-p_{02}\right), \label{eq:s01}\\
    \sigma_{1}=&\frac{3}{4}\mu_{2} U
    -\frac{1}{2}\left(\rho_{1}-\rho_{2}\right)g R^{2}+
    \left(\frac{3}{2}\mu_{2}+3\mu_{1}\right)F_{2}, \label{eq:s11}\\
    \sigma_{n}=&6
    R^{n-2}\left(\frac{\mu_{1}}{2\left(n-1\right)}+\frac{\mu_{2}}{2
      n}\right)F_{n+1} \quad\forall n\geq 2 \label{eq:sn1}.
\end{align}
\nomenclature{$p_{0i}$}{Reference pressure in the $i$'th fluid phase. \nomunit{Pa}}%
Similarly, the shear stress condition (\cref{eq:marangoni}), containing the
Marangoni force, gives the relations 
\begin{align}
    \sigma_{1}=&-\frac{3}{2}\mu_{2} U
    +3\left(\mu_{2}+\mu_{1}\right)F_{2}, \label{eq:s12}\\
    \sigma_{n}=&2
    R^{n-2}\frac{2n-1}{n\left(n+1\right)-6}\left(\mu_{1}+\mu_{2}\right)F_{n+1}
    \quad\forall n\geq 2. \label{eq:sn2}
\end{align}
The simultaneous solution to \cref{eq:s01,eq:s11,eq:sn1} and
\crefrange{eq:s12}{eq:sn2} is given by
\begin{align}
  \sigma_{0}=&\frac{R}{2}\left(p_{01}-p_{02}\right),\\
  \sigma_{1}=&\frac{9}{2}\mu_{1}\left( U- U_{\text{HS}}\right)
  +\frac{1}{2}\mu_{2}\left(6 U-9 U_{\text{HS}}\right),\label{eq:sigma1}\\
  \sigma_{n}=&0\quad\forall n\geq2,\\
  F_{2}=&\frac{3}{2}\left( U- U_{\text{HS}}\right),\\
  F_{n}=&0\quad\forall n\geq3,
\end{align}
where $p_{0i}$ is the reference pressure in the respective phases and
\begin{equation}
  U_{\text{HS}} = \frac{2\Delta\rho|\v{g}| R^2}{9\mu_2},
\end{equation}
where $\Delta\rho=(\rho_1-\rho_2)$, is the Stokes result for the terminal
velocity of a hard sphere. By insertion, one finds that the expression
for $\sigma_{1}$ is zero if $ U$ is replaced by the solution
given by Hadamard and Rybczynski, which is consistent with the
assumption of two clean fluids.
\nomenclature{$U_{\text{HR}}$}{Terminal velocity of a clean drop.\nomunit{m/s}}%
\nomenclature{$U_{\text{HS}}$}{Terminal velocity of a hard sphere.\nomunit{m/s}}%

The resulting expressions for the stream functions can now be given by
\begin{align}
    \Psi_{1}\left(r,\theta\right)&=\frac{3}{4}\left(
     U- U_{\text{HS}}\right)
    \left(r^2-\frac{r^4}{R^2}\right)\sin^{2}\left(\theta\right),\\
    \Psi_{2}\left(r,\theta\right)&=\frac{1}{2}\left(- Ur^{2} 
     +\frac{3}{2} U_{\text{HS}}Rr
     +\left( U-\frac{3}{2} U_{\text{HS}}\right)\frac{R^3}{r}\right)
             \sin^{2}\left(\theta\right). 
\end{align}
One may also notice that the internal stream function is identically
equal to zero if $ U$ is replaced by
$ U_{\text{HS}}$. This shows that if the droplet is falling with
the same velocity as a hard sphere, the Marangoni forces will balance
the shear forces from the external fluid, resulting in a uniform
velocity inside the droplet equal to the droplet velocity.

The above analysis does not give any restrictions on the velocity of
the droplet, other than the demand of keeping the Reynolds number
low. An exotic case would be that of a hovering droplet, meaning that
the Marangoni forces balance the forces induced by gravity. In
the absence of an energy input \eg from a temperature gradient
\citep{young1959} or from an asymmetric release of surfactants
\citep{masoud2014}, both of which are interesting systems in their own right,
a hovering drop will clearly violate the conservation of energy.

\begin{figure}
  \begin{center}
    \subfloat[The vector field $\v{u}$ around the hovering droplet.]{
      \label{fig:lev-drop-vector}
      \includegraphics[width=0.4\linewidth]{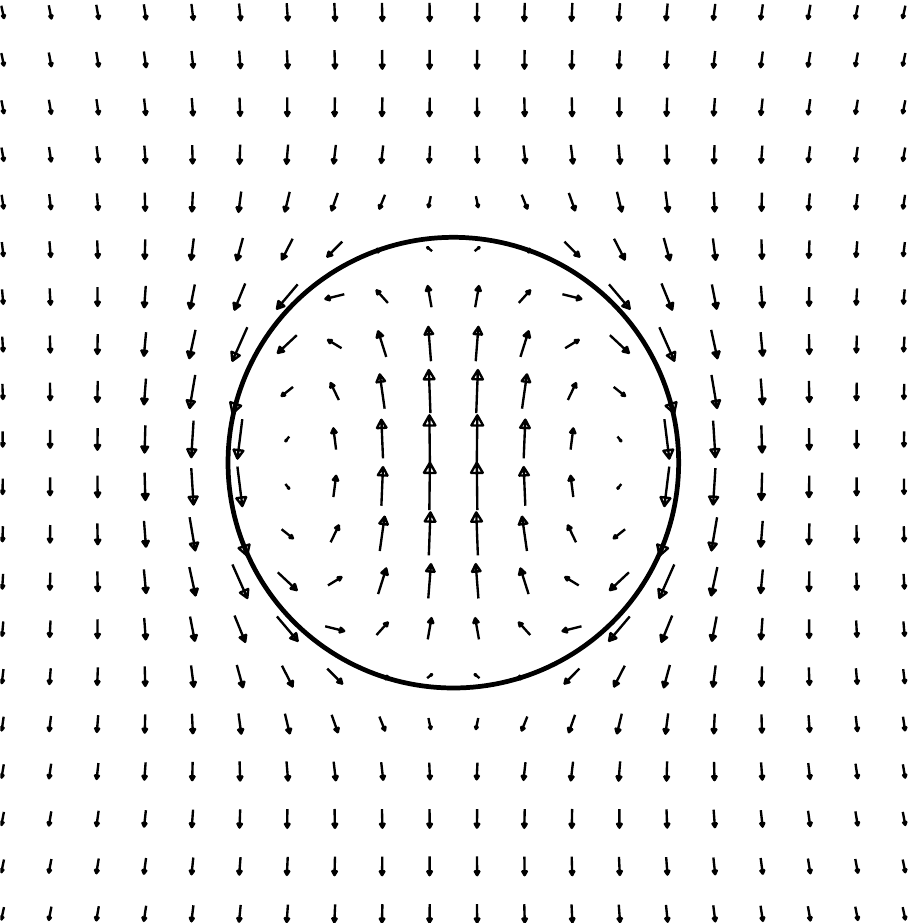}
      }\quad
      \subfloat[Signed magnitude of the velocity along the z- and x-axis, $\theta = 0$ and
      $\theta = \pi/2$ respectively.]{
      \label{fig:lev-drop-z-vel}
      \includegraphics[width=0.5\linewidth]{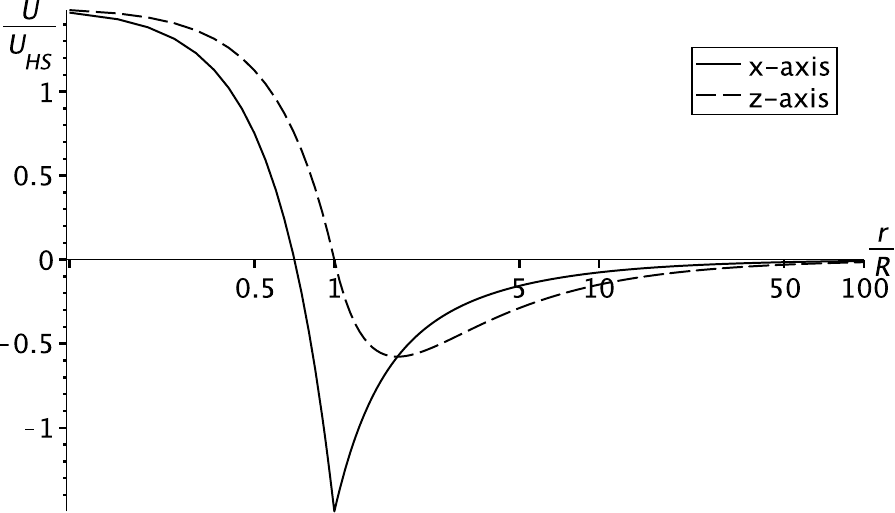}
      }
  \end{center}
  \caption{The velocity field in the case of a hovering drop (terminal velocity
  equal to zero). In (a) the vector field is shown close to the drop. In (b)
line plots of the velocity along the polar (z-axis) and the azimuthal (x-axis)
directions. It is seen that the velocity field decays to zero far away from the
drop. Note that the sharp kink in (b) is caused by the drop interface.}
   \label{fig:lev-drop}
\end{figure}

The expression for the stream function shows that there is a non-zero velocity
field in this case. This is obvious from the fact that in the presence of
Marangoni forces, the viscous stress tensor cannot be zero both inside and
outside the droplet, and hence there must be gradients in the velocity field. In
\cref{fig:lev-drop} we plot (a) the vectors of the velocity field around the hovering
drop and (b) the decay of the velocity field far away from the drop. 
This case corresponds of course to $U=0$, so the coordinate systems of the drop
and the laboratory coincide.

To proceed, one may consider the energy balance in this system, in order to pick
physically acceptable solutions. The energy equation for creeping flow can be
obtained from the Navier-Stokes equations as:
\begin{align}
  \dot{\text{e}}_{\text{K}}=&\frac{\partial}{\partial
    t}\left(\rho\frac{u^2}{2}\right)= 
    \v{u}\cdot\left(\div{\v{T}}\right)+\rho\v{u}\cdot\v{f}_{\text{b}}+ \v{u}\cdot\v{f}_{\text{I}}(r,l)\delta\left(\v{x}-\v{x}_{\text{I}}(r,l)\right) \,\text{d}r\,\text{d}l \\ =&\div{\left(\v{u}\cdot\v{T}\right)}-\v{T}\colon\grad{\v{u}}+\rho\v{u}\cdot\v{f}_{\text{b}}+ \v{u}\cdot\int_\Gamma\v{f}_{\text{I}}(r,l)\delta\left(\v{x}-\v{x}_{\text{I}}(r,l)\right)\, \text{d}r\, \text{d}l,
\label{eq:energy}
\end{align}
\nomenclature{$\text{e}_{\text{K}}$}{The specific kinetic energy. \nomunit{J/m$^3$}}%
\nomenclature{$\v{f}_{\text{I}}$}{Interfacial force. \nomunit{N}}%
\nomenclature{$\delta\left(\v{x}\right)$}{Dirac delta-function. \nomunit{1/m}}%
\nomenclature{$\v{x}_{\text{I}}$}{Material points on the interface. \nomunit{m}}%
where $\delta(\v{x}-\v{x}_\text{I})$ is a Dirac delta-function which is singular
at the interface and $r$ and $l$ is the parametrisation of the interface. The first term on the right of \cref{eq:energy} is the energy flux passing through a fluid
interface (or any finite volume in the general case) and the second term is the energy dissipation in a fluid
element. The third term is the energy provided by the body force term, while the last term is the energy dissipated in the interface due to the action of the surfactants.
For the sake of brevity we will refer to this dissipation as ``energy
consumption'' (or ``energy production'' in the opposite case), even though
energy can of course not be consumed or produced. 

\begin{figure}
  \begin{center}
    \includegraphics[width=0.35\linewidth]{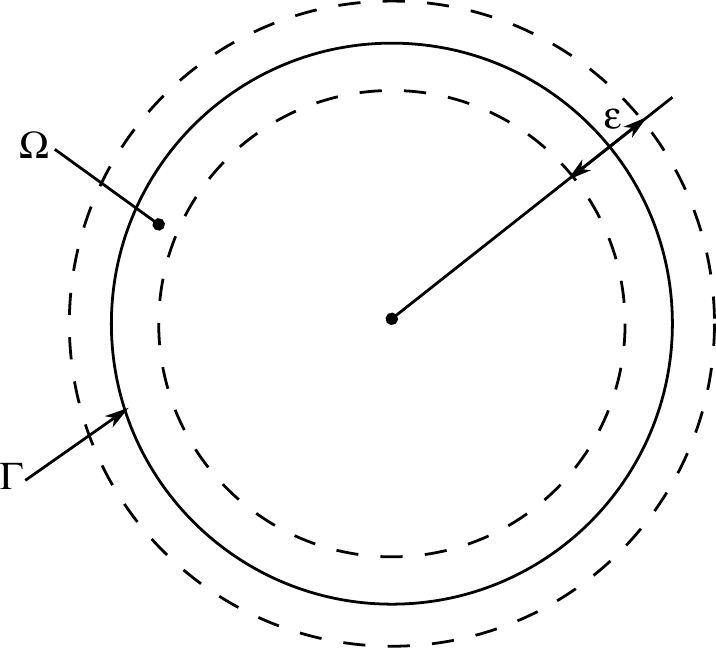}
  \end{center}
  \caption{The integration domain $\Omega$ of thickness $\epsilon$ around the droplet interface $\Gamma$.}
  \label{fig:domain}
\end{figure}
\begin{figure}
  \begin{center}
    \includegraphics[width=0.5\linewidth]{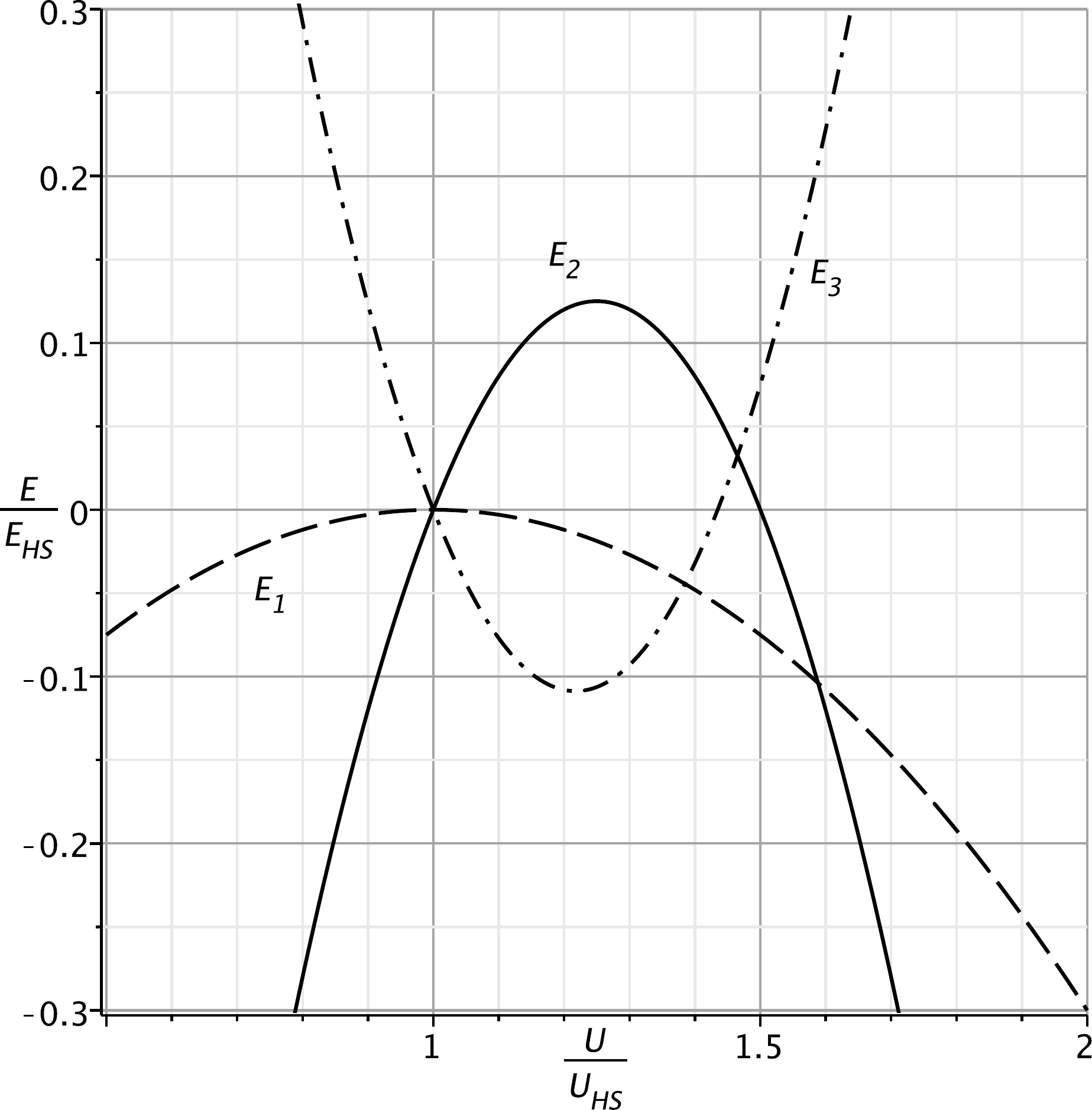}
  \end{center}
  \caption{Energy consumption for the falling droplet. Negative values indicate
    energy consumption, while positive values are energy production. The term
    $E_3$ is the third term on the right-hand side of \cref{eq:energy_balance},
    viz. $E_3 = E_1 - E_2$,
    and we see clearly that the interval in which the interface consumes energy
    is bounded by the Stokes and the Hadamard-Rybczynski terminal velocities,
    cf.\ \cref{eq:energy_bound}. Outside these limits, the interface produces energy, which is unphysical.}
  \label{fig:energy}
\end{figure}

We will now look at the energy balance of the interface itself. This is achieved by integrating \cref{eq:energy} over a volume just enclosing the interface (see \cref{fig:domain}):
\begin{align}
0=\int_\Omega\div{\left(\v{u}\cdot\v{T}\right)}\,\text{d}\v{x} -\int_\Omega\v{T}\colon\grad{\v{u}}\,\text{d}\v{x} +\int_\Omega\rho\v{u}\cdot\v{f}_{\text{b}}\,\text{d}\v{x} +\int_\Omega\v{u}\cdot\int_\Gamma\v{f}_{\text{I}}\delta\left(\v{x}-\v{x}_{\text{I}}\right)\, \text{d}S\,\text{d}\v{x}
\end{align}
After applying Gauss' theorem and letting $\epsilon$ approach zero (following
\citet{hansen2005}), one obtains
\begin{equation}
    0=-\underbrace{ \oint\v{u}_{2}\cdot\v{T}_{2}\cdot\v{n}\, \text{d}S }_{E_1}
  +\underbrace{ \oint\v{u}_{1}\cdot\v{T}_{1}\cdot\v{n}\, \text{d}S }_{E_2}
  + \underbrace{ \oint\left(\v{u}_{1}\cdot\v{t}\right) \left(\v{t}\cdot\grad\sigma\right)\, \text{d}S }_{E_3},
\label{eq:energy_balance}
\end{equation}
where we have labelled the terms for reference in \cref{fig:energy}, and where
\begin{equation}
\v{u}\cdot\v{T}\cdot\v{n}=-p u_r+\mu \left(2 u_r
\pd{u_r}{r}+u_\theta\pd{u_\theta}{r} -\frac{u_{\theta}^{2}}{r}+
\frac{u_\theta}{r}\pd{u_r}{r}\right) 
\end{equation}
in spherical coordinates. We have also used an inward pointing normal vector so
that energy flux into the droplet is positive.  
Since there is no net movement of the drop, the integration over the droplet of
the body force term in \cref{eq:energy} will be zero. Thus
\cref{eq:energy_balance} shows that the energy consumption in the interface together with the energy dissipation in the droplet at stationary conditions must equal the energy flux into the droplet interface from the surrounding fluid side.

The three terms in \cref{eq:energy_balance} are
plotted individually in Figure~\ref{fig:energy}. The figure shows that there is
an interval where the interface is consuming energy to keep the constant velocity. Since we are not providing external energy to the interface, this is the only allowed velocity interval. By setting the third term in \cref{eq:energy_balance} to zero, a second order equation for the velocity gives: 
\begin{equation}
 U= U_{\text{HS}} \quad \vee \quad  U=3\frac{\mu_{1}+\mu_{2}}{3\mu_{1}+2\mu_{2}}
 U_{\text{HS}}\label{eq:energy_bound},
\end{equation}
as the bounding interval. So, the permissible solutions for a viscous sphere falling at
steady state in a gravitational field surrounded by a quiescent liquid
under the influence of Marangoni forces are bounded by the Stokes solution for
the hard sphere and the Hadamard-Rybczynski solution for clean liquids.
Note again the contrast here with the stagnant-cap model, where these two bounds are
assumed \emph{a priori}.
\nomenclature{$E$}{Energy consumption.\nomunit{W}}%
\nomenclature{$E_{\text{HS}}$}{Energy consumption of a hard sphere.\nomunit{W}}%
\nomenclature{$E_{\text{I}}$}{Energy consumption in the interface.\nomunit{W}}%

To compare this result directly to the stagnant-cap model, one may consider the 
terminal velocity of the droplet given as a function of the interfacial tension,
\begin{align}
U=&\frac{3\left(\mu_1+\mu_2\right)}{3\mu_1+2\mu_2}U_{\text{HS}}-\frac{2\sigma_{1}}{9\mu_1+6\mu_2} \label{eq:vel_sig}\\
 =&U_{\text{HR}}-\frac{2\sigma_{1}}{9\mu_1+6\mu_2},
\end{align}
where the reader is reminded that the interfacial tension is
$\sigma=\sigma_{0}-\sigma_{1}\cos\left(\theta\right)$ and that $U_{\text{HR}}$ is the
Hadamard-Rybczynski velocity. This gives the following expression for the drag
force on the droplet:
\begin{equation}\label{eq:sigma_drag}
\text{F}_{\text{D}}=\frac{4\pi \mu_2 UR}{1+\beta}\left[\frac{3}{2}\beta+1+\frac{\sigma_1}{3\mu_2U}\right],
\end{equation}
\nomenclature{$\beta$}{Relative viscosity($\mu_{1}/\mu_{2}$).\nomunit{1}}%
where $\beta$ is the ratio of the inner and outer fluid viscosity. In
comparison, the drag force obtained with the stagnant cap model can be written
as
\begin{equation}\label{eq:scm_drag}
\text{F}_{\text{D}}=\frac{4\pi \mu_2
UR}{1+\beta}\left[\frac{3}{2}\beta+1+f_1\left(f_2^{-1}\left(\frac{1+\beta}{\mu_2U}\sigma_\Delta\right)\right)\right],
\end{equation}
where $\sigma_\Delta$ is the difference between the maximum and minimum value of
interfacial tension as defined in \citet{sadhal1983}, and $f_1$ and $f_2$ are
trigonometric functions of the cap angle \citep{hatanaka1988}. \citet{davis1966}
assumed that $\sigma_\Delta$ was limited by a constant value $\Pi^*$, making 
the argument in $f_2^{-1}$ approach zero when the droplet radius (hence the 
terminal velocity) increases. One then obtains the desired behaviour with the
drag force approaching that for clean droplets.

\subsection{The continuous-interface model}
Proceeding from this result, we will derive a mechanical interface model which links the
interfacial concentration of surfactants to the coefficient of interfacial tension.
To achieve this we will use arguments from molecular considerations, and link the
dynamic equations directly to the Marangoni force.

It is assumed that the surfactant molecules are subjected to a force field, $\v{f}$,
and that their action on each other due to thermal fluctuations is governed by a Wiener
process, \ie a force given by $\lambda\v{F}_s$ where $\lambda$ is a scaling
parameter for the normalised stochastic Wiener function $\v{F}_s$.
It is customary \citep{giona2004} to model the fluid friction on
each molecule by a Stokesian force term, $\v{f}=\psi \v{u}$, where $\psi$ is
a friction constant and $\v{u}$ is
the velocity of the species surrounding the molecule in question. This
leads to Brownian motion, whose stochastic behaviour in the diffusion-controlled
regime is governed by the Langevin equation \citep{giona2004}
\begin{equation}
\d{\v{x}}{t}=-\frac{\v{f}}{\psi}-\frac{\sigma}{\psi}\v{F}_s.
\end{equation}
This equation corresponds to a Fokker-Plank equation which is 
the macroscopic advection diffusion equation \citep{castiglione1999},
\begin{equation}
\pd{\Gamma}{t}+\div{\v{j}}=D_s\grad^2{\Gamma},
\end{equation}
where $\Gamma$ is the interfacial concentration of the species in question, $D_s=1/2(\sigma/\psi)^2$
is the interfacial diffusion coefficient, and $\v{j}=\v{f}/\psi \Gamma$ is the flux. This
advection-diffusion equation is the typical starting point for modelling the 
transport of surfactants on the droplet interface \citep{levich1962,leal2007}.

In the situation where surfactants retard the velocity of the droplet to that of
a hard sphere, the advection-diffusion equation dictates that the only possible
surfactant concentration profile is a constant one. This conclusion is the basis
for the stagnant cap model, namely that one section of the droplet surface has
a constant concentration of surfactants, and the remainder is completely empty
of surfactants. This means that the coefficient of surface tension abruptly 
goes from one constant value on the cap section to another (generally higher)
constant value on the rest of the droplet. A surfactant concentration profile
being piecewise constant is hard to relate to the Marangoni force
\citep{sadhal1983}, which depends on a gradient in the interfacial tension.

In order to relate the Marangoni force to the retardation of covered droplets,
we are forced to alter the advection-diffusion equation approach used \eg by
\citet{levich1962}. Alternatively, as is done here, we may consider a 
force balance on the surfactant layer, \emph{viz.}
\begin{equation}
\Gamma \d{\v{u}}{t}=\div{\v{T}_\text{s}}+\v{f}_\text{M},
\end{equation}   
where $\Gamma$ is the concentration of surfactants, $\v{T}_\text{s}$ is the
two-dimensional surface stress tensor and $\v{f}_\text{M}$ is a body force
affecting the surface layer, namely the Marangoni force. It should be noted that
we now treat the surfactants as a two-dimensional continuum on the interface of
the droplet. Assuming steady-state and inserting the expression for the
Marangoni force, the force balance becomes
\begin{equation}
0=\v{t}\cdot \div{\v{T}_\text{s}}-\v{t}\cdot\grad{\sigma}.
\end{equation}

We now assume that the surfactants form an inviscid continuum, i.e. we neglect
here surface (shear and dilatational) viscosities, so the stress
tensor reduces to only a pressure term, $\v{T}_{s}=-p_\text{I}\v{I}$, giving the
following relation between the pressure in the interfacial layer and the
coefficient of surface tension:
\begin{equation}
p_\text{I}=\alpha-\beta \sigma,
\end{equation} 
where $\alpha$ and $\beta$ are constants to be determined. The pressure in the
interface is related to the concentration of surfactants through a constitutive
relation such as the Langmuir-Blodgett equation of state \citep[see
\eg][]{langmuir1917}. Neglecting the surface viscosities is a reasonable
approximation here, as it is a weaker effect than e.g. surface tension and the 
Marangoni force, but we note that interesting phenomena do arise from the surface viscosity 
\citep{agrawal1979} and may consider the extension to include this in future
work.

The aim of the present approach is to give an alternative model to the stagnant-cap
model. The new model should have some predictive power, explaining some results
obtained in experiments; in particular we will consider the paper by \citet{griffith1962}. In
that paper, relatively low concentrations of surfactant are used, starting from
almost pure fluids. In this case, we will assume that the concentration of
surfactants in the interfacial layer is small, and therefore use a linear
approximation to the Langmuir-Blodgett equation of state, giving a linear
relation between the interfacial tension and the concentration of surfactants
\begin{align}
\Gamma(\theta)&=C-\frac{\sigma}{k}\\
&= C - \frac{\sigma_0}{k} + \frac{\sigma_1}{k}\cos(\theta)\\
&=\Gamma_\text{avg}+\frac{\sigma_1}{k}\cos\left(\theta\right),
\label{eq:ift-surfconc}
\end{align}
where the last line is obtained by integrating $\Gamma(\theta)$ over the sphere and
setting the result equal to $ 4\pi R^2\Gamma_\text{avg}$.

It is readily apparent from \cref{eq:ift-surfconc} that $\Gamma$ as a 
function of $\theta$ is symmetric about $\Gamma_{\text{avg}}$, and that the 
minimum is given by $\Gamma_{\text{avg}}-\sigma_1/k$, which implies that
\begin{equation}
\sigma_1\le k\Gamma_{\text{avg}}.
\end{equation}
Positing that there exists a maximum value for the repulsive force between the
surfactants, the
surfactant would be released from the interface and dissolve into the bulk phase,
\cref{eq:ift-surfconc} shows that $\Gamma$ must be less than some maximum value $\Gamma_{\infty}$. $\Gamma_{\infty}$ is
known as the maximum packing concentration in the surfactant literature.
Inserting $\Gamma_{\infty} \ge \Gamma$ into \cref{eq:ift-surfconc} one obtains 
\begin{equation}
\Gamma_{\text{avg}}\le\Gamma_{\infty}-\sigma_1/k,
\label{eq:avg_max}
\end{equation}
giving
\begin{equation}
\sigma_1\le\frac{k\Gamma_{\infty}}{2}.
\end{equation}

In arriving at these expressions, it was implicitly assumed that there
is no limit to the forces each surfactant molecule can absorb from the surrounding
liquids. In reality, the surfactant molecules will bend and twist if they are
subjected to large stresses. Taking this into account, \ie requiring that
each molecule can at most absorb a force of magnitude $F_{\infty}$, one obtains
a restriction on the Marangoni shear stress $\tau = \sigma_1\sin(\theta)/R$,
namely $\tau/\Gamma\le F_{\infty}$. In addition
to this comes the requirement that $\Gamma\le \Gamma_{\infty}$ as discussed
previously.

Writing it out in full, this expression for the maximum interfacial shear stress is
\begin{align}
\max_{\theta}\frac{\tau}{\Gamma(\theta)}=\max_{\theta}&\frac{\frac{1}{R}\sigma_1\sin(\theta)}
  {\Gamma_{\text{avg}}+\frac{\sigma_1}{k}\cos(\theta)}\le F_{\infty},
\end{align}
or equivalently
\begin{align}
 \max_{\theta}\,\sigma_1\left(\frac{\sin(\theta)}{R}-\frac{F_{\infty}}{k}\cos(\theta)\right)
  \le F_{\infty}\Gamma_{\text{avg}},
\end{align}
giving a restriction on the rate of change of the interfacial tension,
\begin{align}
 \sigma_1\le& \frac{k\Gamma_{\text{avg}}}{\sqrt{1+\left(\frac{k}{F_{\infty}R}\right)^2}}\le k\Gamma_{\text{avg}},\label{eq:sig_gamma_avg}
\end{align}
which upon insertion of \cref{eq:avg_max} yields
\begin{equation}
\sigma_1\le \frac{k\Gamma_{\infty}}{1+\sqrt{1+\left(\frac{k}{F_{\infty}R}\right)^2}}\le \frac{k\Gamma_{\infty}}{2}.
\label{eq:anequation}
\end{equation}
Using \cref{eq:vel_sig} to eliminate $\sigma_1$ from \cref{eq:sig_gamma_avg},
one obtains now an expression for the lowest terminal velocity allowed for
a drop of radius $R$. Normalising this by the terminal velocity of a hard sphere,
$\chi=U/U_{\text{HS}}$, gives
\begin{align}
\chi\geq \frac{3\left(\mu_1+\mu_2\right)}{3\mu_1+2\mu_2}
-&\frac{2}{9\mu_1+6\mu_2}\frac{k\Gamma_{\text{avg}}}{U_{\text{HS}}\sqrt{1+\left(\frac{k}{F_{\infty}R}\right)^2}}.
\end{align}
At this point it is convenient to introduce the viscosity ratio $\beta
= \mu_1/\mu_2$, as well as the quantity $R_c=k/F_{\infty}$.
This critical radius $R_c$ is the largest droplet radius such that the forces from the
surfactants on the liquid are large enough to retard the 
droplet to the Stokes terminal velocity ($\chi=1$). We denote the drop radius
normalised by the critical radius as $x=R/R_c$. Using the fact that this
expression should become 1 at $x=1$,
\ie that drops with the critical radius fall like hard spheres, we
obtain $\Gamma_{\text{avg}}/R_c=\sqrt{2}\Delta\rho g/3F_{\infty}$, and
the previous equation simplifies to
\begin{align}
\chi\geq\frac{3\left(\beta+1\right)}{3\beta+2}-&\frac{\sqrt{2}}{3\beta+2}\frac{x^{-2}}{\sqrt{1+x^{-2}}}.
\end{align}

Notice that in the expression for $\Gamma_\text{avg}/R_c$ preceding this
equation, the maximum force a surfactant molecule
can absorb, $F_\infty$, is a material constant for the surfactant. This means
that for a given surfactant, the average interfacial concentration is
directly proportional to the critical radius. Thus, in low
surfactant-concentration experiments, one may use the critical radius as
a measure of equilibrium interfacial concentration of surfactants,
$\Gamma_\text{avg}$. The reader is reminded that $\Gamma_\text{avg}$ can 
be related to the bulk concentration of surfactants, $C$, by a Langmuir isotherm
\begin{equation}
  \Gamma_\text{avg} = \Gamma_\infty \frac{aC}{1+aC}
  \label{eq:langmuir}
\end{equation}
where $a$ is a constant.

Recalling the proof in \cref{sec:sphere_drop} that a drop
cannot fall slower than a hard sphere of equal radius, since this violates the
conservation of energy, we obtain the final expression for the relative velocity
$\chi$ as
\begin{equation}\label{eq:rel_vel_almost_fin}
  \chi(x) \geq
  \begin{cases}
     1 & \text{if}\; x\leq 1\rm{,}\\
    \frac{3\left(\beta+1\right)}{3\beta+2}-
    \frac{\sqrt{2}}{3\beta+2}\frac{x^{-2}}{\sqrt{1+x^{-2}}}  & \text{if}\; x> 1. 
  \end{cases}
\end{equation}
Notice that this expression is continuous at $x=1$, but the derivative is
discontinuous at this point, cf.\ \cref{fig:griff_fit}. Notice also that 
when $R \gg R_c$, \ie $x^{-2} \ll 1$, the expression for $x>1$ approaches the
Hadamard-Rybczynski result. Thus the inequality must be replaced by equality in
the $x \gg 1$ limit. In the $x \to 1$ limit, equality is also required since it
is observed that the drops fall like hard spheres. The simplest expression which
is correct in both these limits is obtained by replacing the inequality with
equality for the entire expression, \emph{viz.}
\begin{equation}\label{eq:rel_vel_fin}
  \chi(x) =
  \begin{cases}
     1 & \text{if}\; x\leq 1\rm{,}\\
    \frac{3\left(\beta+1\right)}{3\beta+2}-
    \frac{\sqrt{2}}{3\beta+2}\frac{x^{-2}}{\sqrt{1+x^{-2}}}  & \text{if}\; x> 1. 
  \end{cases}
\end{equation}
This is the prediction of the continuous-interface model for the transition in
terminal velocity as a function of drop radius.

\section{Discussion}
\label{sec:discussion}
In 1953, \citet{savic1953} introduced the stagnant cap model (SCM) as an
explanation of the experimental results obtained by \citet{bond1927,bond1928}.
The SCM incorporates the effect of surfactants through a rigid cap where
a no-slip boundary condition is used. It is obvious that this will lead
to a terminal velocity of droplets bounded by the Stokes velocity and the
Hadamard-Rybczynski velocity. In later works several iterations of the SCM have been
proposed\citep{griffith1962, davis1966, harper1973,sadhal1983}, and presently 
two different versions exist, namely the model proposed by \citet{griffith1962} and
the model proposed by \citet{davis1966}. The \citeauthor{griffith1962} approach
is based on calculating the cap angle from
a criterion based on the average interfacial pressure difference, while
\citeauthor{davis1966} employ a local criterion based on the capillary tension
and the interfacial shear forces.

\citet{hatanaka1988} give a review of the experiments performed by
\citet{bond1928} and \citet{griffith1962} and compare the two versions of the
SCM with the experimental results. \citeauthor{hatanaka1988} show that the model of
\citeauthor{davis1966} gives better agreement with the experiments performed by
\citet{bond1928}, while the \citeauthor{griffith1962} model gives better
agreement with the experimental results performed by \citeauthor{griffith1962}
himself. It appears that the difference between the experimental results by 
\citeauthor{griffith1962} and those of \citeauthor{bond1928} 
is too large to be governed by the same mechanism. Note here that while the
experiments due to \citet{bond1928} use fluids which are assumed to be pure,
in the experiments by \citet{griffith1962} a surfactant is deliberately added 
at known bulk concentrations. In general, one considers the experiments performed by
\citet{griffith1962} to be more reliable, since the experimental setup there is better
controlled, taking advantage of developments in our understanding of
chemistry and fluid mechanics, as well as developments in experimental
equipment, not available at the time of \citet{bond1928}.

\begin{figure}
  \begin{center}
    \includegraphics[width=5.8cm]{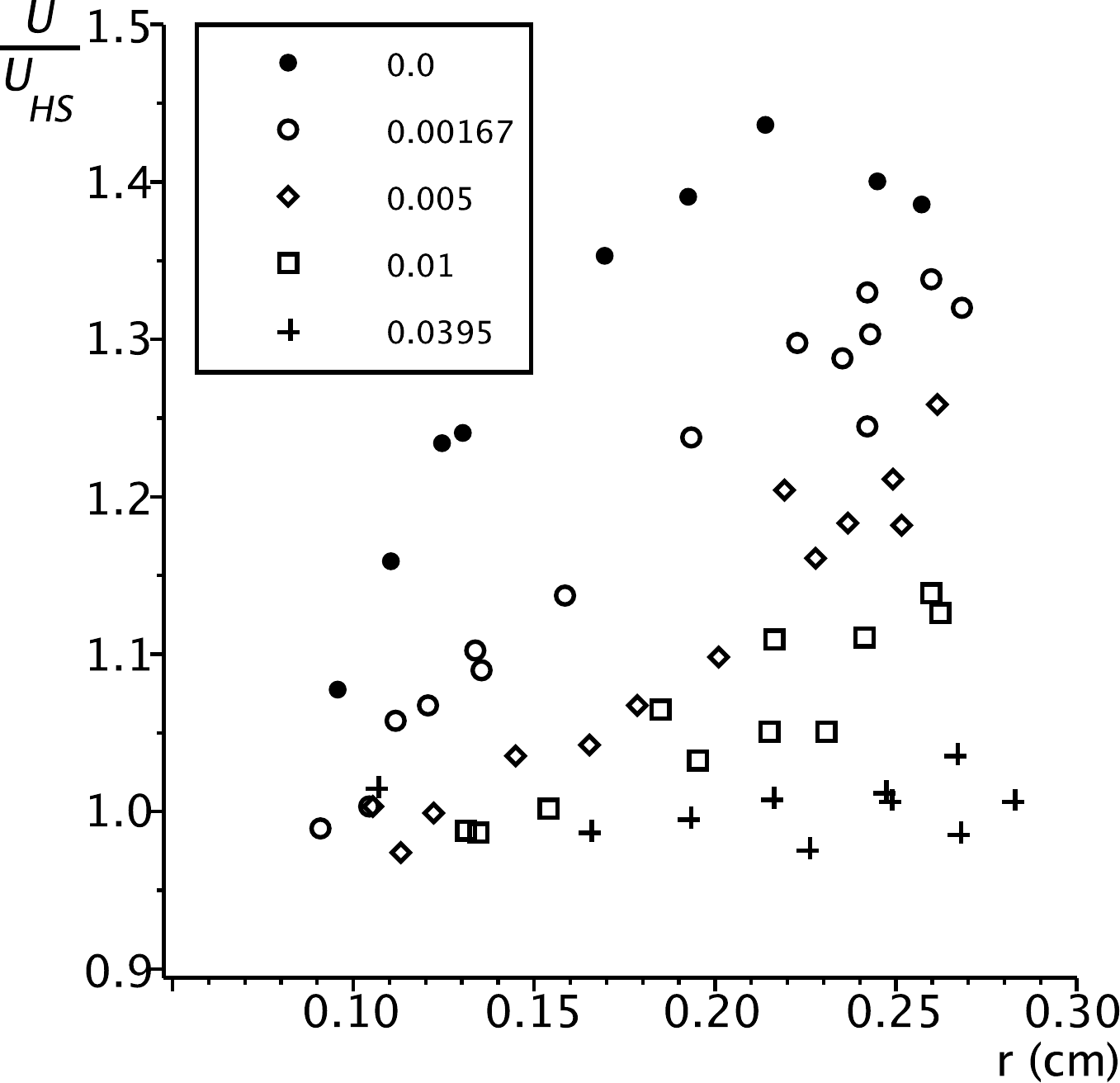}
  \end{center}
  \caption{Droplet terminal velocities for different drop sizes and at different
  surfactants concentrations (\citet[Fig. 9]{griffith1962}). The concentration
is given in $\unit{kg/m^3}$ and the highest concentration corresponds to $35.4
\unit{ppm}$.}
  \label{fig:griff_raw}
\end{figure}

\cref{fig:griff_raw} shows one set of experiments performed by
\citet{griffith1962}, extracted from Figure 9 in his work. 
The figure shows the results from an experiment with
drops of ethylene glycol with Aerosol 61 surfactant falling in a reservoir of a mineral
oil, using different concentrations of surfactant. The results are
plotted as the relative velocity $\chi = U/U_{HS}$ versus the drop radius. 
Notice that the base fluids employed by Griffith are not free of surface active agents,
indicated by the results without any added surfactant showing the same trend of
approaching the hard sphere terminal velocity as the radius decreases.

\begin{figure}
  \begin{center}
    \includegraphics[width=5.8cm]{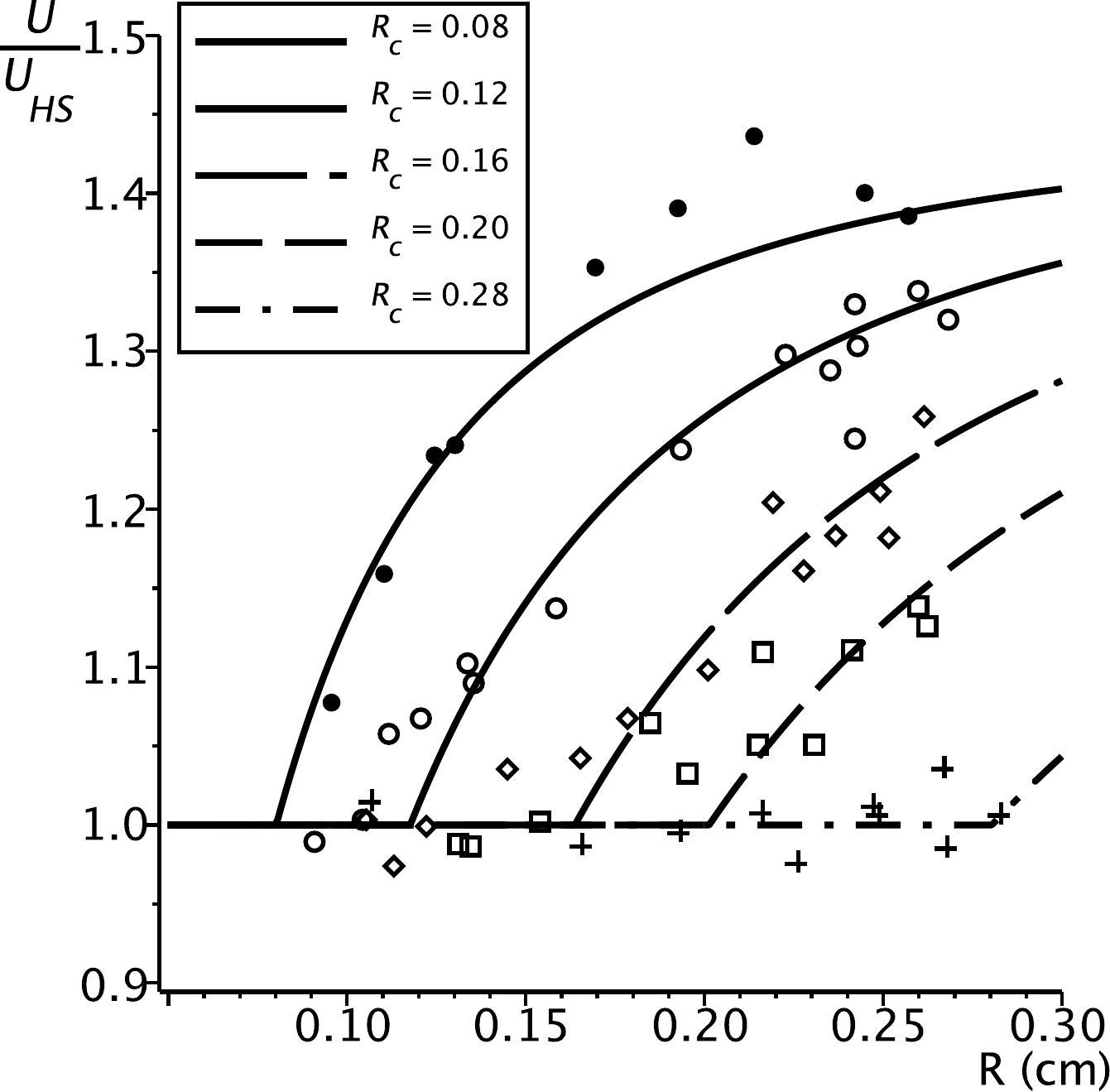}
  \end{center}
  \caption{Nonlinear curve fit of \cref{eq:rel_vel_fin} to the raw data by \citet{griffith1962}.}
  \label{fig:griff_fit}
\end{figure}

By fitting \cref{eq:rel_vel_fin} to these data points, we may calculate the
critical radius $R_c$ below which $\chi = 1$ in the experiments performed by
Griffith. This is shown in \cref{fig:griff_fit}, where the obtained values of
$R_c$ for each of the five concentrations is shown in the legend. Note that in
theory, if perfectly pure fluids were used, $R_c \to 0$ as $C \to 0$. This is
not the case for these experiments.

\begin{figure}
  \begin{center}
    \includegraphics[width=5.8cm]{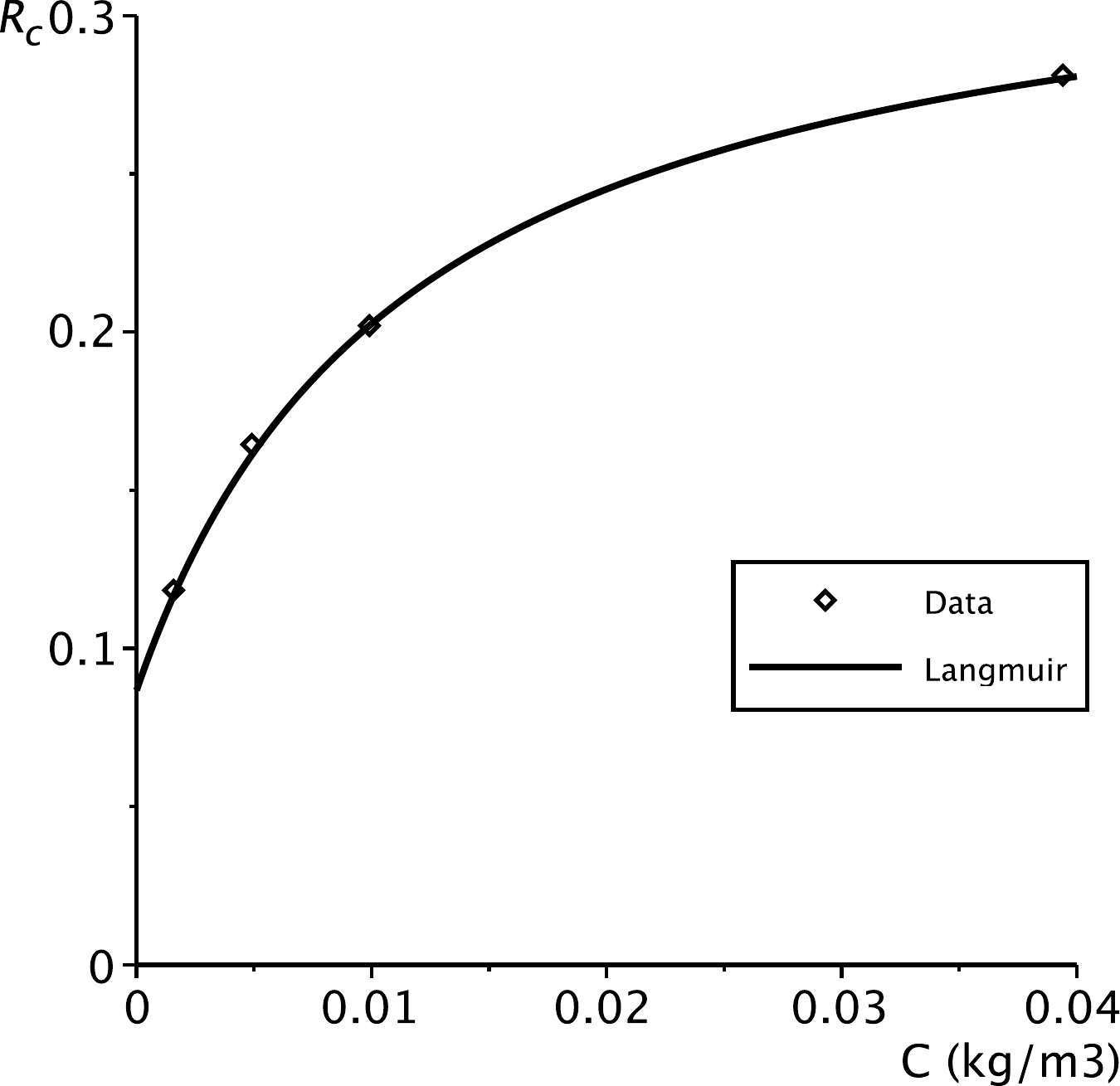}
  \end{center}
  \caption{$R_c$ obtained from the fitting of \cref{eq:rel_vel_fin} to the
  experimental data, plotted against bulk surfactant concentration, together
with a Langmuir isotherm fitted to these points.}
  \label{fig:griff_langmuir}
\end{figure}

As outlined in the previous section, the proposed continuous-interface model
predicts that the critical radius is directly proportional to the interfacial
surfactant concentration, which can again be related to the bulk concentration
through the Langmuir isotherm \cref{eq:langmuir}. This means the critical radius
is also related to the bulk concentration via a Langmuir isotherm, \ie we can
write $R_c = R_c(C)$. Notice,
however, that the isotherm must be modified to account for the fact that
surfactants are still present in the system at $C=0$, \ie $R_c(C=0) \ne 0$. By
replacing the concentration in \cref{eq:langmuir} with $C' = C + C_\text{base}$,
we obtain an isotherm with two unknown parameters, $a$ and $C_\text{base}$. 
Fitting this to the experimental data, as shown in \cref{fig:griff_langmuir}, it
is seen that the critical radii all collapse to the obtained Langmuir isotherm.
This confirms the prediction made by the proposed continuous-interface model.

To complete this discussion, we wish to point out that the stagnant cap model
is likely a good model when the surface-active agents interact like hard particles.
To see this, consider the explanation of the stagnant cap model in
terms of the Marangoni force, as attempted by \citet{sadhal1983}. If the
Marangoni stress balance is valid, then arguably the normal stress balance 
should also be satisfied. In the SCM, none of the
above stress balances are used as boundary conditions. It is then only possible
(given the uniqueness of solutions to the Stokes equation)
to satisfy  both stress balances if one of them can be written in terms of the
boundary conditions used. This is obviously not possible, and hence the normal
stress balance cannot be satisfied.

It is therefore natural to conclude that the
SCM may be applicable in situations where interfacially active components
interact like particles and form a solid cap. In the continuous-interface model
presented here, both stress balances are used in the boundary conditions and
therefore, in contrast
to the SCM, it applies when the interfacially active components are 
amphiphilic molecules which produce a Marangoni force.
\footnote{If the SCM is a reasonable model for interfacially active
components in the form of particles, one might want to re-examine experimental evidence for 
the SCM that is obtained using particle-based flow visualisation
methods, such as the canonical flow visualization photograph due to \citet{savic1953} reproduced in
\cref{fig:savic} here.}

\section{Concluding remarks}
In this paper we have derived the exact solution to the flow inside and around
a circular drop falling at low Reynolds number, with an arbitrarily varying
interfacial tension. By avoiding the use of a surfactant advection-diffusion
equation at the interface, we are able to obtain analytical solutions to the
flow, which has not been possible in previous works. 
We demonstrate that when all the interfacial stress
conditions are taken into account, one obtains a range of simultaneous solutions
for the variation in interfacial tension and for the flow field, including
exotic solutions such as the hovering drop. By appealing to conservation
of energy, we restrict the allowed interval of solutions, and show that the terminal
velocity of a falling drop must lie between the clean drop 
(Hadamard-Rybczynski) and the rigid sphere (Stokes) results.

To proceed with this approach, we propose a new model for how surfactants behave
at the interface of a falling drop. Previous work has assumed the existence of
a stagnant cap of surfactants on the top of a falling drop. In the present
model we do not impose a specific surfactant distribution, but we introduce
a simple model, called the continuous-interface model, which takes into account
the force balance for surfactant molecules at the interface.
It is demonstrated that the model gives a transition in terminal velocity as a function of drop
radius that is consistent with experimental results. Moreover, by fitting the
model to experimental results, we extract values for the critical radius as a
function of bulk surfactant concentration. The model predicts that these should
be related by a Langmuir isotherm, and indeed this is found to be true.
We postulate that our model
is more reasonable for fluid-like surfactant molecules, while the stagnant cap
model may be appropriate for colloidal particles acting as surfactants. Future
work may attempt to identify this difference experimentally.

Ending on a historical note, we have read with interest the relatively recent 
paper by \citet{hager2012} about the life and work of Wilfrid Noel
Bond, who, amongst other achievements, was the first person to observe and
discuss the transition in terminal velocity that we aim to explain with our model. Bond's
untimely demise was surely a great loss not only for his family, but also for
the field of fluid mechanics research.
\footnote{Those with an interest in history may also want to read the similar
exposition on Wolfgang von Ohnesorge by \citet{mckinley2011}.}
\nocite{mckinley2011}

The authors are grateful for stimulating discussions on these matters with  Dr. 
Svend Tollak Munkejord and Professor Bernhard Müller, as well as for the comments
of the anonymous referees which helped improve this paper.
This work was funded by the project \emph{Fundamental understanding of
electrocoalescence in heavy crude oils} coordinated by SINTEF Energy Research.
The authors acknowledge the support from the Petromaks programme of the Research
Council of Norway (206976), Petrobras, Statoil and Sulzer Chemtech.

\bibliographystyle{abbrvnat}
\setcitestyle{authoryear,open={(},close={)}}
\bibliography{references}

\end{document}